# An Energy-Efficient Localization Strategy for Smartphones

Haifeng Liu, Feng Xia[*], Zhuo Yang, Yang Cao

School of Software, Dalian University of Technology,
Dalian 116620, China
f.xia@ieee.org

**Abstract.** In recent years, smartphones have become prevalent. Much attention is being paid to developing and making use of mobile applications that require position information. The Global Positioning System (GPS) is a very popular localization technique used by these applications because of its high accuracy. However, GPS incurs an unacceptable energy consumption, which severely limits the use of smartphones and reduces the battery lifetime. Then an urgent requirement for these applications is a localization strategy that not only provides enough accurate position information to meet users' need but also consumes less energy. In this paper, we present an energy-efficient localization strategy for smartphone applications. On one hand, it can dynamically estimate the next localization time point to avoid unnecessary localization operations. On the other hand, it can also automatically select the energy-optimal localization method. We evaluate the strategy through a series of simulations. Experimental results show that it can significantly reduce the localization energy consumption of smartphones while ensuring a good satisfaction degree.

## 1. Introduction

With smartphones being increasingly pervasive in the past few years, lots of Location-Based Applications (LBAs) have been developed to meet the needs of human life. Those currently popular LBAs include mobile social networking, mobile search, healthcare, traffic monitoring [1, 2, 3], etc.

The core enabler of LBA is the localization technique, which is used to obtain position information. A great number of wireless localization techniques have been outlined by Liu et al. [4], among which GPS is preferred to other techniques such as WiFi and GSM based positioning system, because of its higher accuracy. However, GPS is extremely energy-hungry, which will largely shorten battery lifetime of smartphones when GPS based LBAs are running. This particular theory has been validated by tremendous experimental results [5, 6, 7]. Meanwhile, how to get position information with minimal energy has become one of the primary issues for recent research studies.

[*] Corresponding author: Feng Xia (f.xia@ieee.org)

Numerous solutions have been proposed to address the problem of high localization energy consumption of GPS. They are mainly classified into three categories. The first category is to replace GPS with other methods which are more energy-efficient but less accurate, such as WiFi [8, 9] and GSM [10] positioning system. The second category is to avoid unnecessary GPS localization operations. For instance, it requires no localization operation when the user is stationary or the user does not move beyond the limited range during a particular period. The third category is to adopt the hybrid method which continuously switches among multiple localization techniques.

In this paper, we present an energy-efficient localization strategy for smartphone applications based on the following three observations. First, LBAs do not always require the highest accuracy and their accuracy requirements will vary as the user moves in general. For example, location-based social networking application can change accuracy requirement depending upon the positions of users. Proximity can be efficiently perceived using method proposed by Kupper and Treu [11]. This method calculates the required accuracy limit for each user by utilizing the distance between them. The accuracy limits might range from 10 meters, if the users are close, to several kilometers if they are far apart. Since the application accuracy requirement is relaxed as the user moves, a less accurate but more energy-efficient localization method can be selected. Second, it is always not necessary to locate the user again if it has been away for some time while not beyond the accuracy limited range. Hence, during the particular period, it needs not to locate the user and a waste of energy is avoided. Third, most of the time, it is easy to obtain the velocity of a user. We have observed that the accelerometer is often used with the method proposed by Brezmes et al. [12, 13] to detect whether a user is stationary or not. At this time, we can leverage accelerometer to determine velocity cheaply. In addition, we can also obtain velocity by interacting with intelligent public transport system when we are on a bus.

Based on the above observations, our proposed scheme is designed to save energy in two main respects. On one hand, it can reduce localization operations by dynamically estimating the localization time point. On the other hand, it can select the energy-optimal localization method.

This paper makes the following contributions:

1) We present an energy-efficient localization strategy for smartphone applications which minimizes the energy consumption while satisfying the application accuracy requirement.

2) We evaluate the effectiveness of the proposed strategy through simulations. Experimental results show that our scheme significantly outperforms the existing GPS method.

The remainder of the paper is organized as follows: Section 2 presents the related work, which is followed by the introduction of our energy-efficient localization strategy in Section 3. In Section 4, we evaluate the performance of the proposed strategy. Section 5 concludes the paper.

## 2. Related Work

Recently, with the ever-increasing quantities of LBAs of smartphones, GPS has been frequently used in various applications because it is highly accurate. However, GPS is exceedingly energy-hungry. This will largely shorten battery lifetime of smartphones when LBAs are running.

To solve this problem, many previous works have already studied alternative localization methods to replace the GPS technique. Otsason et al [14] presented a GSM-based indoor localization system. WiFi positioning systems were proposed by Khan and Akidi [15]. Additionally, a great number of other techniques such as Bluetooth [16, 17, 18], UWB [19, 20], and RFID [21, 22], are also used to determine the position information. Nevertheless, most of those methods are less accurate.

Some other works focus on how to improve GPS and combine GPS with other methods to conserve energy and prolong battery life. Zhuang et al [23] presented a localization framework for smartphone applications, which also overviewed energy-efficient localization techniques. Kjargaard [24] summarized many energy-saving methods for smartphones from different levels. Abdesslem et al. [7, 25, 26] used an accelerometer to detect whether the user is stationary or not. If the user has been stationary for a long time, GPS is not necessarily used during this period. These works effectively avoided unnecessary localization operations when the user is in static state. Paek et al [6] presented a rate-adaptive GPS-based positioning system. Compared to duty-cycled GPS, this work dynamically estimated the user's velocity depending on the location-time history of the user. In fact, it leveraged user's daily living habits. Kjaergaad et al [26] used the user's velocity measured by GPS to estimate the localization time point, which is more inaccurate compared with our velocity sampling algorithm. Farrel et al [27] considered to reduce GPS energy consumption by assuming that the localization operation is only required when the user enters a specified area. Constandache et al [28] added consideration for given energy budget to the problem of localization. Lin et al [29] presented a Bayesian estimation framework to model users' locations and sensor errors. However, the energy-optimal localization method is not always selected.

## 3. Energy-Efficient Localization

### 3.1. Overview

We consider LBAs of smartphones equipped with GPS, WiFi, and GSM positioning interfaces. As the user moves, LBAs need to continually determine and update positions. When the user is stationary, we can use the method proposed by Huang et al [25, 26, 27] to detect the static state and reduce unnecessary localization energy consumption. Hence, we only consider energy-efficient localization when the user is always in mobile state.

Application accuracy requirements vary as the user moves. We assume that we can obtain the application accuracy requirement anytime, and let $a_t$ be the application accuracy requirement at time $t$. If there are multiple applications at the same time, $a_t$ will be equal to the minimum value of accuracy at time $t$.

Further, for GPS, WiFi, and GSM positioning systems, we make the following assumptions. The energy consumption per localization operation for the three systems is a constant value, which is denoted with $E_g$, $E_w$, and $E_m$ respectively. Additionally, the localization accuracy is also constant, respectively denoted with $a_g$, $a_w$, and $a_m$. Moreover, the process of localization is assumed to occur instantaneously.

As for the user's movement in a period of time, let $E_T$ denote the total energy consumption for duration $T$, and $E_{t_i}$ denote the energy consumption at time $t_i$, which is the $i$ th localization time point. $E_T$ can be defined as follows.

$$E_T = E_{t_1} + E_{t_2} + E_{t_3} + \cdots + E_{t_i} + \cdots \quad i = 1,2,3,\ldots \quad (1)$$

In order to minimize the total energy consumption $E_T$, we propose an energy-efficient localization strategy, whose flow chart is depicted in Figure 1.

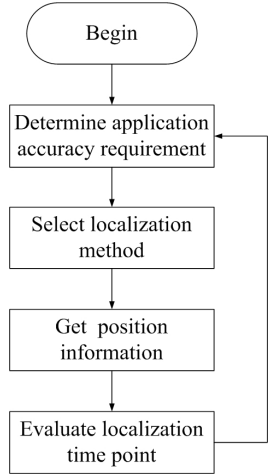

**Fig. 1.** Flow chart of our energy-efficient localization strategy

Our solution considers conserving energy consumption in the following two respects. First, we avoid unnecessary localization operations by dynamically estimating the next localization time point and sampling velocity. If the user has been in movement for some time within the range of application accuracy limit, it is not necessary to locate the user. Second, when it is time to locate the user again, we use the energy-optimal method by calculating the average energy consumption of each strategy and selecting the least one.

## 3.2. Estimating Localization Time Point

To avoid unnecessary localization operations, we sample velocity to compute the movement range of the user and dynamically estimate the localization time point. It shall be the time to locate the user again as long as the user moves beyond the application accuracy limited range.

It is assumed that we can obtain the velocity of a user anytime because of the observation that it is easy to determine velocity when an accelerometer is used to detect whether the user is stationary or not. Furthermore, it can also be based on other observations. For example, we can use the bus's velocity as a user's velocity by interacting with intelligent public transport system when the user is on a bus, or use GPS to determine velocity when it is being used to get position information. In addition, we let $v(t_{ij})$ denote the sampled velocity at time $t_{ij}$, where $i$ and $j$ are defined as the $i$ th time in localization and the $j$ th time to sample velocity respectively.

The detailed process of estimating localization time point is described as follows.

---
**Algorithm 1.** Estimating localization time point

1. At localization time point $t_{ij}$, ser $r_i = 0$;
2. Compute $v_e(t_{ij}) = \alpha \times v(t_{ij}) + (1-\alpha) \times v_e(t_{(i-1)(j-1)})$;
3. Obtain $a_{t_i}$, select $M_{t_i}$ using Equation (4);
4. Compute $t_s = (a_{t_i} - a_{M_{t_i}})/v_e(t_{ij})$;
5. Set $j = j+1$, $t_{ij} = t_{i(j-1)} + t_s \times \beta$ $(0 < \beta \leq 1)$;
6. Compute $v_e(t_{ij}) = \alpha \times v_e(t_{ij}) + (1-\alpha) \times v_e(t_{i(j-1)})$;
7. Compute $r_i = r_i + v_e(t_{ij}) \times t_s \times \beta$ $(0 < \beta \leq 1)$;
8. **if** $ri < (a_{t_i} - a_{M_{t_i}})$ **then**
9.     go to step 5;
10. **else**
11.     **return** $t_{ij}$;
12. **end if**
---

In the algorithm, $M_{t_i}$ is defined as the method selected at localization time point $t_i$. It can be obtained from Equation (4), which will be described in the following subsection. $v_e(t_{ij})$ denotes the velocity estimated by the Exponentially Weighted Moving Average (EWMA) method, during the interval between two consecutive velocity samplings. $r_i$, which represents the estimated movement distance since the localization time point $t_{ij}$, can be obtained from multiplying $v_e(t_{ij})$ and the particular interval. The coefficient $\alpha$ is a factor ranging from 0 to 1. A higher value of $\alpha$ means a closer velocity estimated to the current velocity. Additionally, $\beta$ is a factor used to adapt the time point of the next velocity sampling. When the coarse localization interval $t_s$ is computed, we sample velocity after $t_s \times \beta$ and compute velocity estimated by EWMA method to determine movement distance. By using this method, we can dynamically estimate the localization time point.

Also, parallel to the process, we need to monitor the application accuracy requirement. Whenever it changes, we must go back to step 1 to locate the user again.

### 3.3. Selecting Localization Method

The objective of this subsection is to select the energy-optimal localization method, which satisfies the application accuracy requirement when it is time to locate the user again. Some researchers [29] directly select the method that consumes the lowest energy. However, it is not always the best. Generally, the lower energy a method consumes, the less accurate it will be. For example, it is known that GPS is highly accurate compared with other localization methods, while its energy consumption is extremely high. If it is used to locate a user, it will take shorter time for the user to move beyond the application limited range. In contrast, it will take longer time. Let $E_{low}$ denote the energy consumed by a low energy-consuming method, and $E_{high}$ for a high energy-consuming method, and their time interval between two consecutive localization operations can be denoted with $T_{low}$ and $T_{high}$ respectively. If two applications using the two methods run at the same time, the total energy consumptions, $E^1_{low}$, and $E^2_{high}$ can be calculated as follows.

$$E^1_{low} = \frac{E_{low}}{T_{low}} \times T \qquad (2)$$

$$E^2_{high} = \frac{E_{high}}{T_{high}} \times T \qquad (3)$$

It is obvious that if $E_{high}/T_{high} < E_{low}/T_{low}$, then $E^2_{high} < E^1_{low}$. Therefore, based on the above theory, we present our selection method using Equation (4), where GPS, WiFi, GSM positioning methods are respectively represented by symbols *g*, *w*, and *m*.

$$M_t = \arg\min_{g,w,m} \left\{ \frac{E_{M_t}}{(a_t - a_{M_t})/v_e(t)} \right\} \qquad (4)$$

$M_t$ is selected from the set of localization methods {*g,w,m*} at time *t*, which is the method that produces the minimum ratio of $E/T$.

## 4. Performance Evaluation

In this section, we provide a simulation study of the energy-efficient strategy to assess the performance of the proposed algorithm. The following two metrics are considered in our simulations.
- Total energy consumption: This is the total energy consumed by LBAs during the period that application is running.
- Satisfaction degree: This is the ratio of the time that the evaluated location satisfies the application accuracy requirement to the total time that application is running.

### 4.1. Settings

To evaluate the performance of our strategy, the following mobile pattern is considered.

The time unit is defined as one second, and the minimum and maximum velocity are set to 1*m/s* and 10*m/s* respectively for a mobile user. Let *v(t)* denote the velocity of an user at time *t*. We further assume that the acceleration changes every $T_1$ seconds, detailed in Equation (5).

$$if\ t \bmod T_1 = 0\ then\ a(t) = \begin{cases} 0\ or\ 1 & if\ v(t\text{-}1) = 1m/s \\ -1\ or\ 0 & if\ v(t\text{-}1) = 10m/s \\ -1,\ 0\ or\ 1 & otherwise \end{cases} \quad (5)$$

where *a(t)* denotes the acceleration at time *t*. When *t* is equal to the integer times of $T_1$, a new acceleration is generated at once. When *v(t-1)* is equal to the minimum velocity 1*m/s*, the acceleration is equal to either 0 or 1. Likewise, when *v(t-1)* is equal to the maximum velocity 10*m/s*, the acceleration is equal to either 0 or -1. In other cases, the acceleration is equal to one of the set {-1,0,1} randomly. The velocity *v(t)* is computed by Equation (6).

$$v(t) = \begin{cases} v(t-1) + a(t) & if\ t \bmod T_1 = 0 \\ v(t-1) & if\ t \bmod T_1 \neq 0 \end{cases} \quad (6)$$

The simulation runs 3600s. The application accuracy requirement is changed every 600s as 500m, 300m, 150m, 120m, 80m, and 50m. For GPS, WiFi, and GSM, using the data from previous work [26], accuracy is 10m, 50m, and 150m, and energy consumption is 1425mJ, 545mJ, and 20mJ, respectively. The acceleration generates again every 3s, using Equation (5).

### 4.2. Result Analysis

Using the mobile pattern above, we evaluated our strategy, as shown in Figures 2, 3, 4, and 5.

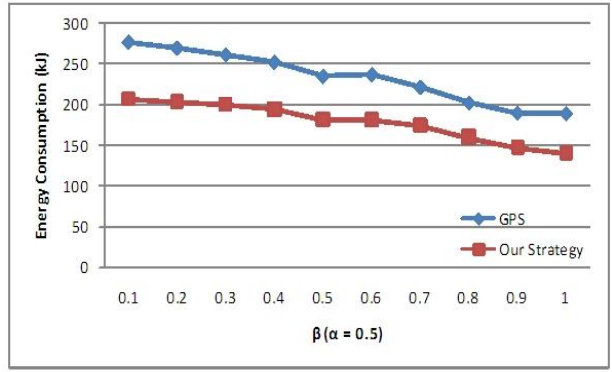

**Fig. 2.** Energy Consumption of GPS versus our strategy (*α*=0.5)

Figures 2 and 3 depict the total energy consumption and the satisfaction degree of our strategy and GPS, respectively, when *α* used in EWMA method is set to 0.5 and the value of *β* ranges from 0.1 to 1.0. We observed that as *β* increases, the total energy consumption is reduced, as shown in Figure 2, but the satisfaction degree is also lowered, as shown in Figure 3, for both our strategy and GPS. The reason is that, with the increasing value of *β*, the interval of velocity sampling will also increase, then the localization operation after the user is beyond the application accuracy limit occurs repeatedly. In other words, the interval of two consecutive localizations is prolonged so that the frequency of localization operation is reduced, which leads to reduce localization energy consumption. At the same time, the satisfaction degree is lowered for the same reason. These results prove the effectiveness of our estimation method of the localization time point.

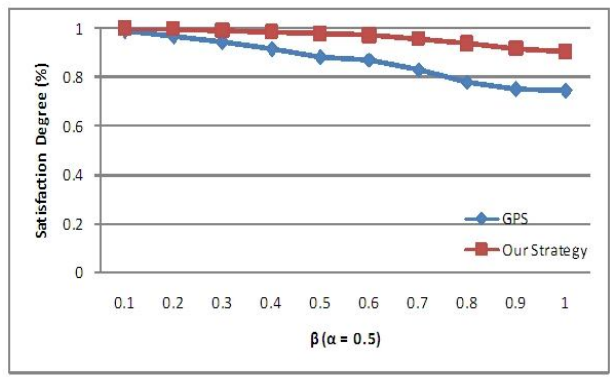

**Fig. 3.** Satisfaction Degree of GPS versus our strategy (α=0.5)

Compared with GPS, our energy-efficient localization strategy conserves more energy. The reason is that our strategy is designed to select the energy-optimal method. Commonly, the selected method is less accurate than GPS, but the interval of two consecutive localizations is shorter than the one based on GPS. As a result, the time point we estimate to locate the user again is more accurate and the corresponding satisfaction degree is better. This is proved by the simulation results.

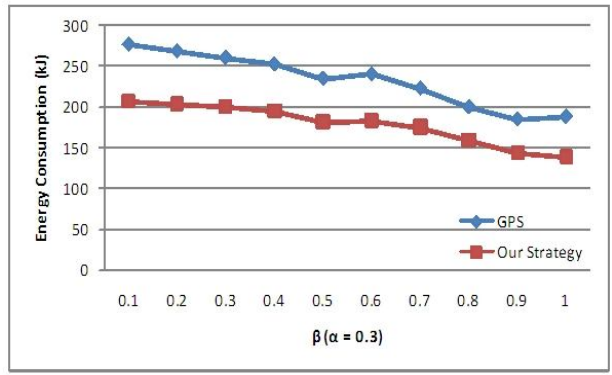

**Fig. 4.** Energy Consumption of GPS versus our strategy (α=0.3)

Figures 4 and 5 show the simulation results when *α* is set to 0.3 and the value of *β* ranges from 0.1 to 1.0. They also proved the effectiveness of our strategy.

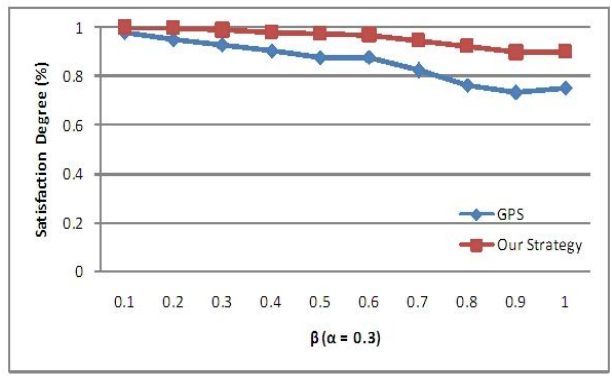

**Fig. 5.** Satisfaction Degree of GPS versus our strategy (α=0.3)

As shown in the figures above, we found that the simulation results seem not to be affected by the value of *α*. This is because that the acceleration in our simulation is not changed significantly. In fact, the smaller *α* will be better when the acceleration increases as the user moves, because the previous velocity is more useful to estimate movement distance. Otherwise, the larger *α* will be better. For example, assuming the time interval is one second, and there are four velocities {2,4,8,16} in four intervals. Therefore, the acceleration is {0,2,4,8}, implying that the acceleration becomes larger and larger. The moving distance is 2+4+8+16=30. We sample two velocities at the starting time and the ending time, which are 1*m/s* and 4*m/s*. The estimated moving distance is (2×0.5+16×0.5)×4=36 when $\alpha$ is equal to 0.5, (2×0.7+16×0.3)×4=24.8 when $\alpha$ is equal to 0.3, and (2×0.9+16×0.1)×4=13.6 when $\alpha$ is equal to 0.1. As $\alpha$ is set to an appropriate smaller value, the estimated velocity and moving distance are more effective and closer to the real values.

## 5. Conclusions

In this paper, we presented an energy-efficient localization strategy for smartphone applications. It is designed to conserve energy mainly in the two respects below. First, it dynamically determines the next localization time point and sample velocity. This is intended to reduce energy consumption by avoiding unnecessary localization operations. Second, it selects the energy-optimal method to help reduce energy consumption when the estimated localization time point comes. We have evaluated the performance of the proposed strategy through a series of simulations. The results show that it can significantly reduce energy consumption and improve satisfaction degree. One of our future works is to apply some other advanced techniques, Bluetooth for example, to achieve further improvement. Another is to consider how to collaborate on localization with users in the vicinity.

## Acknowledgement

This work was partially supported by the Natural Science Foundation of China under Grant No. 60903153, Zhejiang Provincial Natural Science Foundation of China under Grant No. Y108685, the Fundamental Research Funds for Central Universities (DUT10ZD110), the SRF for ROCS, SEM, and DUT Graduate School (JP201006).